\documentclass [11pt,a4paper,twoside] {article}

\usepackage[dvips]{graphicx}
\usepackage[english]{babel}
\usepackage{times}
\usepackage{multicol}
\usepackage{amsmath}
\usepackage{amssymb}
\usepackage{amsfonts}
\usepackage{fancyhdr}
\usepackage{exscale}

\makeatletter
\renewcommand{\@seccntformat }[1]{\csname the#1\endcsname\quad}
\renewcommand\section{\@startsection {section}{1}{0mm}{-3.5ex \@plus -1ex \@minus -.2ex}{2.3ex \@plus.2ex}{\normalfont\normalsize\bfseries}}
\renewcommand\subsection{\@startsection{subsection}{2}{0mm}{-3.25ex\@plus -1ex \@minus -.2ex}{1.5ex \@plus .2ex}{\normalfont\normalsize\textit}}
\makeatother

\begin{document}
\vspace*{-25mm}
\begin{minipage}[t][10mm][c]{170mm}
\begin{center}
\begin{footnotesize}
Proceedings of The 1$^\text{st}$ International Conference on \textbf{D}iffusion in
\textbf{S}olids and \textbf{L}iquids\\DSL-2005, July 6-8, 2005, University of Aveiro, Aveiro, Portugal
\end{footnotesize}
\end{center}
\end{minipage}
\vspace*{3mm}

\begin{center}

{\Large
REACTION ENHANCED DIFFUSION IN SPHERICAL MEMBRANES}\\
\vspace*{5mm}
{\large%
Serge SHPYRKO$^{1,2*}$, Vladimir M.SYSOEV$^{1}$, }\\
\vspace*{5mm}
\begin{footnotesize}
$^1$ Kiev National University, Physical Faculty, pr.Glushkova,6, Kiev, Ukraine
\\
$^2$Institute for Nuclear Research, Ukrainian Academy of Sciences,
pr.Nauki,47, Kiev Ukraine\\
\vspace*{3mm}
*Corresponding author.\; Email: serge\_shp(\#)yahoo.com\\
\end{footnotesize}
\end{center}

\vspace*{2mm}
\hrule
\vspace*{2mm}

\textbf{Abstract}\\

\begin{small}
\hspace*{5mm}
The reversible reactions like $A+B \leftrightharpoons C$ in the
many-component diffusive system affect the diffusive properties of the
constituents. The effective conjugation of irreversible processes of
different dimensionality takes place due to the stationarity in the system
and can lead to essential increase of the resulting diffusive fluxes.
The exact equations for the spatial concentration profiles of the
components are difficult to treat analytically. We solve
approximately the equations for the concentration profiles of the
reaction-diffusion components in the spherical geometry in the
application to the problem of the enhanced oxygen transfer through a
biological membrane and to the mathematically similar problem of
surface diffusion in a solid body. In the latter case the spherical
geometry can be an adequate tool for describing the surface of a real
solid body which can be modeled as a fractal object formed of sequences
of spherical surfaces with different radii.

\end{small}
\hrule
\vspace*{10mm}
\normalsize
\begin{multicols}{2}
%#############################################################################
% Please note that "floats" such the "table" and "figure" environments are not
% allowed inside the "multicols" environment
%#############################################################################

\section{Introduction}
\vspace*{-3mm}

\hspace*{5mm}
The phenomenon of "facilitated through chemical reaction diffusion" is
familiar to various domains of science. Its essence consists in the diffusion enhancement through the intermittance of some chemical reaction involving the diffusive components. A straightfoirward example is the so called "vacancy enhanced diffusion" of doped impurity into a solid body. Since long ago it was widely known that the presence of crystal imperfections is able to facilitate the impurity penetrating thus considerably enlarging the effective diffusion coefficient. From the thermodynamical point of view the facilitation comes from the presence of several fluxes of the diffusant (as if it were propagating through several different channels, or rather consist of several species having different diffusivity) and intervenience of the {\it reversible} chemical reaction between them. Thus, in a solid body the impurity atioms are known to reside either in the nodes of the crystalline lattice, (substitutional uimpurity) or in the space between nodes (interstitial impurity), the latter being fast diffusants and the former - slow ones. The reversible interchange of these two species tooks place and is facilitated by the presence of vacansies and eigen interstitial atoms of the matrix.
An another example is taken from the biology domain and it concerns the phenomenon of the facilitated oxygen transfer through cellular membranes \cite{Murray}.
The principal reaction scheme consists in reversible "tying" the
ligande molecules by some slow macromolecular carrier.  The facilitated transport of oxygen is possible via some
fermentative kinetics (which should not however affect the
chemical properties of $O_2$) and especially through reversible reaction with
haemoglobine or myoglobine. In what follows we refer for concreteness to this example  though our consideration can be applied
to the vast variety of situations falling into the same reaction
scheme.

The reaction process with $Hb$ and oxygen can be (although very
schematically) represented by the following expression

\vspace{-0.2cm}
\begin{equation}
O_2+ Hb  \rightleftharpoons   HbO_2
\end{equation}
which means the formation of an (unstable) complex $HbO_2$; the
rates of forward and reverse reactions are $k_1$ and $k_-$.

The system of
balance equations for three constituents can be written as follows:

\begin{eqnarray}
  D\Delta c = \rho + q  \nonumber \\
  D_p \Delta c_p = \rho \\
  D_p \Delta c_c = -\rho \nonumber
\end{eqnarray}

where $c$, $c_p$ and $c_c$ stand for concentrations of $O_2$, $Hb$
and complex $HbO_2$ respectively; $D$ and $D_p$ are diffusion
coefficients for $O_2$ and $Hb$ (assuming that both $Hb$ and
$HbO$-complex possess the same diffusion coefficients since the
$Hb$ molecule is much larger in size than oxygen one and both pure
$Hb$ and the complex $HbO_2$ must have similar diffusive
properties). Instead $Hb$ some other protein, e.g.  $Mb$ can
appear. The consuming term $q$ (here set to $0$) should be added,
for example, if
treating the problem of muscular oxygen transport with myoglobine.
The reaction rate $\rho$ from elementary kinetics reads as follows

\begin{equation}
\rho = k_1cc_p - k_-c_c
\end{equation}
imposing minimal nonlinearity on the system.

\section{Reaction-Diffusion System in Spherical Geometry}

\vspace*{-3mm}

\hspace*{5mm}
We start with considering the problem (2)-(3) within the spheroidal
shell of internal and external radii $a$ and $b$. The consideration of this problem within plain and cylindric geometry was performed not long ago \cite{Murray} and was intended to describe the problem of oxygen saturation in the muscular tissue. Our choise of the spherical form of the membranes, besides purely mathematical interest, is motivated by several reasons. Besides the
interest in  elucidating the biological problem of facilitated oxygen
transport in the (spherically shaped) alveols we refer also to the
above mentioned diffusion problem
in crystalline bodies. Usually the mathematical models of impurity
diffusion use the plain geometry as  a tool to represent the boundary
of a solid body as the platform for the diffusion in a bulk.  But the surface
of a crystal is by no means plain, and its inhomogeneities can affect
the diffusion effects (intuitively it is clear from the fact that merely
the effective diffusive surface is bigger than the mathematical surface
of the body). So, plain model for the boundary is an approximation which
can be improved. Namely, it is possible to model the surface as a highly
irregular sequence of spheres of different radii, perhaps, forming fractal
structure. Therefore the consideration of the spherical geometry (on a
single sphere) is believed to yield some improved approximation than
just considering the straight plane.

So in the following we speak on the biological problem of oxygen transport
and consider the spherical membrane.
We assume the spherical symmetry so only one coordinate, namely $r$
resides. As boundary conditions for the problem it is natural to take
$c=c_a$ at $r=a$ and $c=c_b$ at $r=b$ and the zero flux of other
components $dc_p/dr = dc_p/dr =0$.
But strictly speaking, the boundary conditions for all three
constituents in real membranes {\em can not } be specified basing on
a set of biological measurements \cite{Murray}. As it can be
demonstrated,
the specification of one or other type of boundaries does not affect
cardinally the shape of the solution across the whole width of the
membrane exept thin layers at the edges. In any case at chosen way of
handling the problem (see the next chapter) the set of boundary
conditions should be
included in the solution by some kind of self-matching procedure.

Integrating the sum of two last equations in (2) yields
\begin{equation}
c_p+c_c=const \equiv K
\end{equation}
(the conservation of the protein content). If we introduce the new
function
\begin{equation}
 Y\equiv c_c/K
\end{equation}
which has a meaning of the ferment saturation function the system of
equations is cast as:
\begin{eqnarray}
D\cdot\frac{1}{r^2}\frac{\partial}{\partial r}
(r^2\frac{\partial c}{\partial r}) = \rho  \\
K\cdot D_p\cdot\frac{1}{r^2}\frac{\partial}{\partial r}
(r^2\frac{\partial Y}{\partial r}) = -\rho
\end{eqnarray}
with
\begin{equation}
\rho= k_1Kc-Y\cdot K(k_1c+k_-)
\end{equation}

Keeping in mind everything said about the boundary conditions
specifications we now try to handle the problem setting
\begin{equation}
  \begin{array}{ll}
       c_{|a}=c_a &  c_{|b}=c_b  \\
       Y_{|a}=Y_a &  Y_{|b}=Y_b
 \end{array}
\end{equation}

where $c_b$ and both $Y_a,\, Y_b$ should be further determined basing
upon restrictions imposed by the biological sense of the problem.

Using the boundary conditions add eqs (6-7) and integrate twice thus
obtaining an expression relating $c(r)$ and $Y(r)$:
\begin{eqnarray}
D\,( c_b -c(r))+D_p\,K\,(Y_b-Y(r))= \nonumber \\
\frac{b-r}{br}\cdot\frac{ab}{b-a}\cdot [D(c_b-c_a)+D_p\,K\,(Y_b-Y_a)]
\end{eqnarray}

\iffalse
Thus the equation for $c(r)$ reads as
follows:
\begin{equation}
D\frac{1}{r^2}\frac{\partial}{\partial r }
(r^2\frac{\partial c}{\partial r})=
k_1K\,c - (k_1 \, c +k_-)\,Y\,K
\end{equation}
where $Y$ should be expressed from (10)
\fi

\section{External Solutions and Facilitated Transport}
\vspace*{-3mm}

\hspace*{5mm}
Expressing $Y$ from (10) and inserting it into (6) and (8) following
equation results:

\begin{center}
\begin{eqnarray}
D\frac{1}{r^2}\frac{\partial}{\partial r}(r^2\frac{\partial
c}{\partial r}) =-\frac{k_-}{D_p}\cdot \left[Dc_b+Y_bKD_p\right]+ \nonumber \\
c\cdot \left\{ \frac{k_-D}{D_p}-
k_1K [Y_b-1 +\frac{D}{D_pK}c_b]\right\}  -
\label{main}  \\
\frac{b-r}{br}
\frac{ab}{b-a} \left\{ \frac{k_-}{D_p}[D(c_a-c_b)+
D_pK(Y_a-Y_b)] - \right.  \nonumber  \\
\left.
\frac{k_1\cdot c}{D_p} [D(c_a-c_b)+D_pK(Y_a-Y_b)] \right\}
+ \frac{k_1D}{D_p}c^2  \nonumber
\end{eqnarray}
\end{center}

The values $Y_b,\,Y_a$ and $c_b,\, c_a$ as said above should be
determined from self-matching conditions.

The equation (\ref{main}) is extremely difficult (if not impossible)
to solve analytically.  Instead of looking for its exact solution
we notice that
(\ref{main}) can be cast (after
scaling variables) as the equation in the form:  $$ \varepsilon
\Delta c = f(c,r),$$ and $\varepsilon$ appears to be a
small parameter ($\sim D/k_1$). This parameter is smaller if the
reaction is more intensive; for biological issues its value varies
within $10^{-4}-10^{-6}$ \cite{Murray}, which allows approximate
perturbative treatment of the problem. Since this parameter enters the
equation near the term of the highest derivative, the problem
appears to be that of the singular perturbation theory. Dropping
out the $\varepsilon$ term one gets so-called "external solution"
to the problem. In general this equation of smaller order
can't satisfy the boundary conditions
and need "suturing" with the
internal (exact) solution around the boundaries. The external
solutions may suit for practical purposes within the bulk of the
body, and would yield the desired
solution of (\ref{main}) provided that the constants (like
$Y_b,\,Y_a,\,c_b$ etc) are chosen properly, namely by means of the
self-matching procedure.
 In this
case (\ref{main}) turns out to be an ordinary algebraic equation, whose
coefficients depend on the values of
$Y_b,\,Y_a,\,c_b,\,c_a$. But in fact these latter themselves are
solutions to the equation at the boundaries. To ensure that this is
the case one must set following self-matching conditions:

\begin{equation}
Y_b=\frac{k_1c_b}{k_1c_b+k_-} \, , \qquad
Y_a=\frac{k_1c_a}{k_1c_a+k_-}
\end{equation}

(These expressions could be easily obtained from (8) setting $\rho=0$
at the boundaries).

The variation of the saturation function and the total oxygen flux are:

\begin{equation} Y_a-Y_b\equiv s =
\frac{k_1k_-(c_a-c_b)}{(k_1c_a+k_-)(k_1c_b+k_-)}
\label{oxyg:flux:1}
\end{equation}
and
\begin{eqnarray}
F(r) \equiv  -D\frac{\partial c}{\partial r}-K\cdot
D_p\frac{\partial Y}{\partial r}
\equiv  \nonumber \\
\equiv F_d+F_f= \frac{1}{r^2}\frac{ab}{b-a}\cdot
\left\lbrack D(c_a-c_b)+ \right.
\label{oxyg:flux:sol}   \\
\left.  D_p\cdot K \cdot
\frac{k_1k_-(c_a-c_b)}{(k_1c_a+k_-)(k_1c_b+k_-)}\right]  \nonumber
\end{eqnarray}
which is the sum of two terms, first of which, $F_d$ is the ordinary
diffusive flux, and the second one, $F_f$ is the additional flux
caused by the reaction.

The self-matching conditions yield two relations between four
values $c_a$, $c_b$, $Y_a$ and $Y_b$ (or equally $s$). From the
point of view of experimentalists these values are not treatable
in the same fashion. Depending on whether we consider the
biological problem of facilitated oxygen flux or mathematically
similar problem of impurity diffusion, the set of
good-to-operate values changes. For the oxygen
transfer we can operate only the value  $c_a$ within wide range of
magnitudes, and measure $s$, but this latter only for extremely
big $c_a$, where it is known to tend to a constant $\bar{s}$ $(\sim
0.6-0.8)$ \cite{Murray}.  And thus in order to get fluxes and
concentrations within moderate $c_a$ values one recurses to an
interpolation which is expected to yield semi-quantitative
results. Let us express $c_b$ from  (\ref{oxyg:flux:1}):

\begin{equation}
 c_b=\frac{k_{-1}c_a(1-s)-
 \frac{k_{-}^2}{k_1}s}{k_1c_as+k_{-}(1+s)} \,.
\label{oxyg:cb}
\end{equation}

 This expression is exact. Hovewer the flux $s$ is itself a
function of $c_a$ to be determined (only its value $\bar{s}$ at
$c_a \to\infty$ is available), so let us go to the
semiqualitative interpolation:
\begin{equation}
c_b\simeq \frac{k_{-}c_a(1-\bar{s})}{k_1c_a\bar{s}+
k_{-}(1+\bar{s})} \,.
\label{oxyg:cb:interpol}
\end{equation}

The expression (\ref{oxyg:cb:interpol}): a) satisfies the limiting
condition $c_b \to 0$ at $c_a=0$; b) at big
 $c_a$ ($k_1c_a\gg k_{-}$) is consistent with
(\ref{oxyg:cb}). For intermediate $c_a$ this is believed to be
a plausible interpolation. So from (\ref{oxyg:cb:interpol}) the
approximate expression for saturation function is
$s\equiv Y_a-Y_b$:

\begin{equation}
s\simeq \frac{k_1c_a\bar{s}}{(k_1c_a+k_{-})}\cdot
\left[ \frac{k_1c_a+2k_{-}}{k_1c_a+k_{-}(1+\bar{s})}\right]
\label{oxyg:satur:interpol}
\end{equation}

From (\ref{oxyg:cb:interpol})-(\ref{oxyg:satur:interpol}),
substituting it into (\ref{oxyg:flux:sol}), we get the
expression for the complete flux through a membrane. The flux at
the external ($r=b$) layer equals

\begin{eqnarray}
F= F_d + F_f \simeq \nonumber \\ \simeq 4\pi \frac{a}{b}\frac{1}{b-a}\cdot
c_a\bar{s}
\left(k_1c_a+2k_-\right)\times
\label{oxyg:flux:interpol} \\
\left[\frac{D}{k_1c_a\bar{s}+k_-(1+\bar{s})} + \nonumber \right. \\
\left.
+\frac{D_pKk_1}{(k_1c_a+k_-)(k_1c_a+k_-(1+\bar{s}))}\right]
\nonumber
\end{eqnarray}
and is schematically depicted on Fig.1 where the ordinary
diffusion flux and facilitated one are shown as functions of
$c_a$.

\begin{center}
{\includegraphics[scale=0.6]{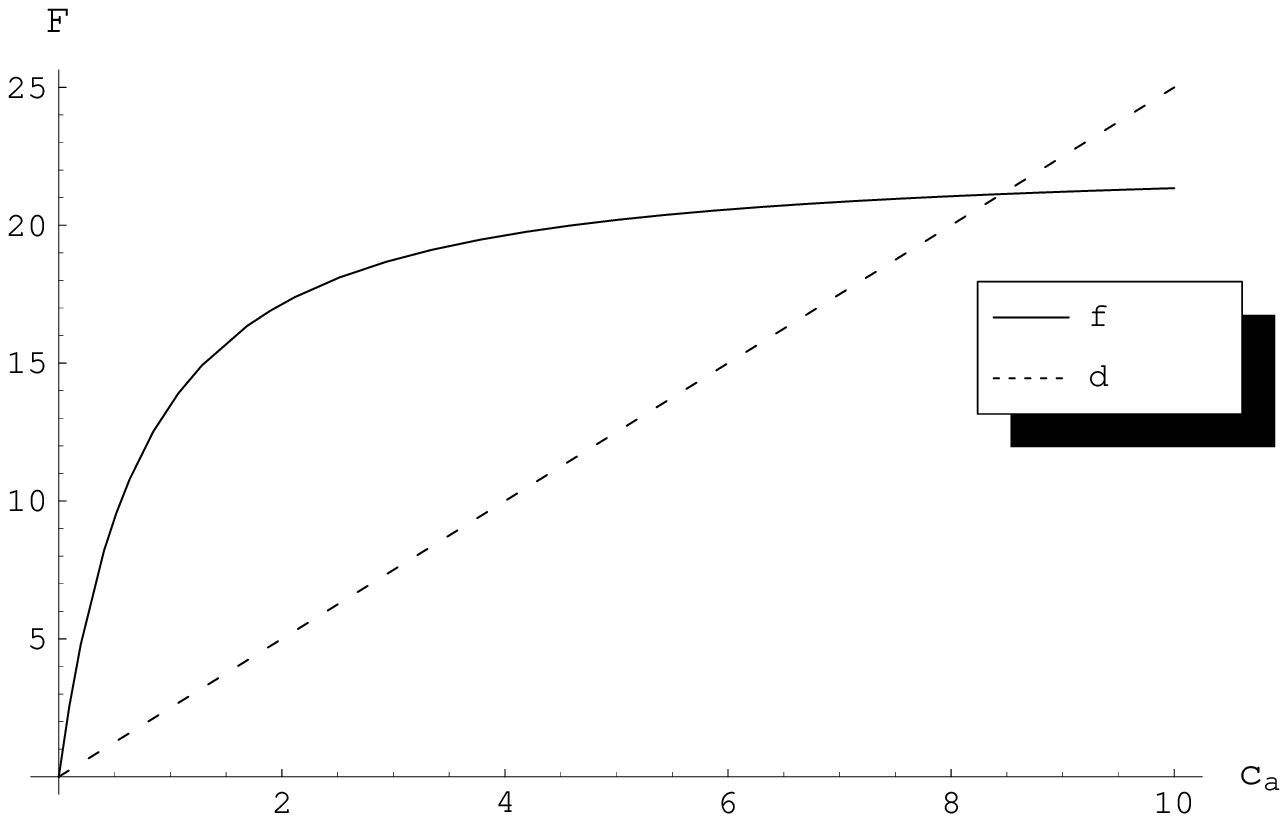}}
Figure 1: Diffusion and reaction facilitated oxygen flux as
function of $c_a$ (arbitrary units). \\
\end{center}

It is interesting to note different character of two fluxes from
the latter formula and from Fig.1: at big $c_a$ the diffusion flux
grows linearly, but the flux component due to
selfconjugation with reaction tends to a constant value. For both
biological problem, and for the problem of vacansion diffusion
enhancement this is clear intuitively: since the amount of carrier
is limited at big values of $c_a$ the substrate simply gets
saturated. As to the moderate or small $c_a$ the relation of two
fluxes is
$$
\frac{F_f}{F_d}=\frac{D_pKk_1}{D k_-}\,.$$

The value $D\gg D_p$ (since $D$ stands for "fast species"), but
under condition $k_1 \gg k_-$ which is true far from chemical
equilibrium, and if it is the case, the facilitated transport
dominates over diffusion.

\begin{center}
{\includegraphics[scale=0.6]{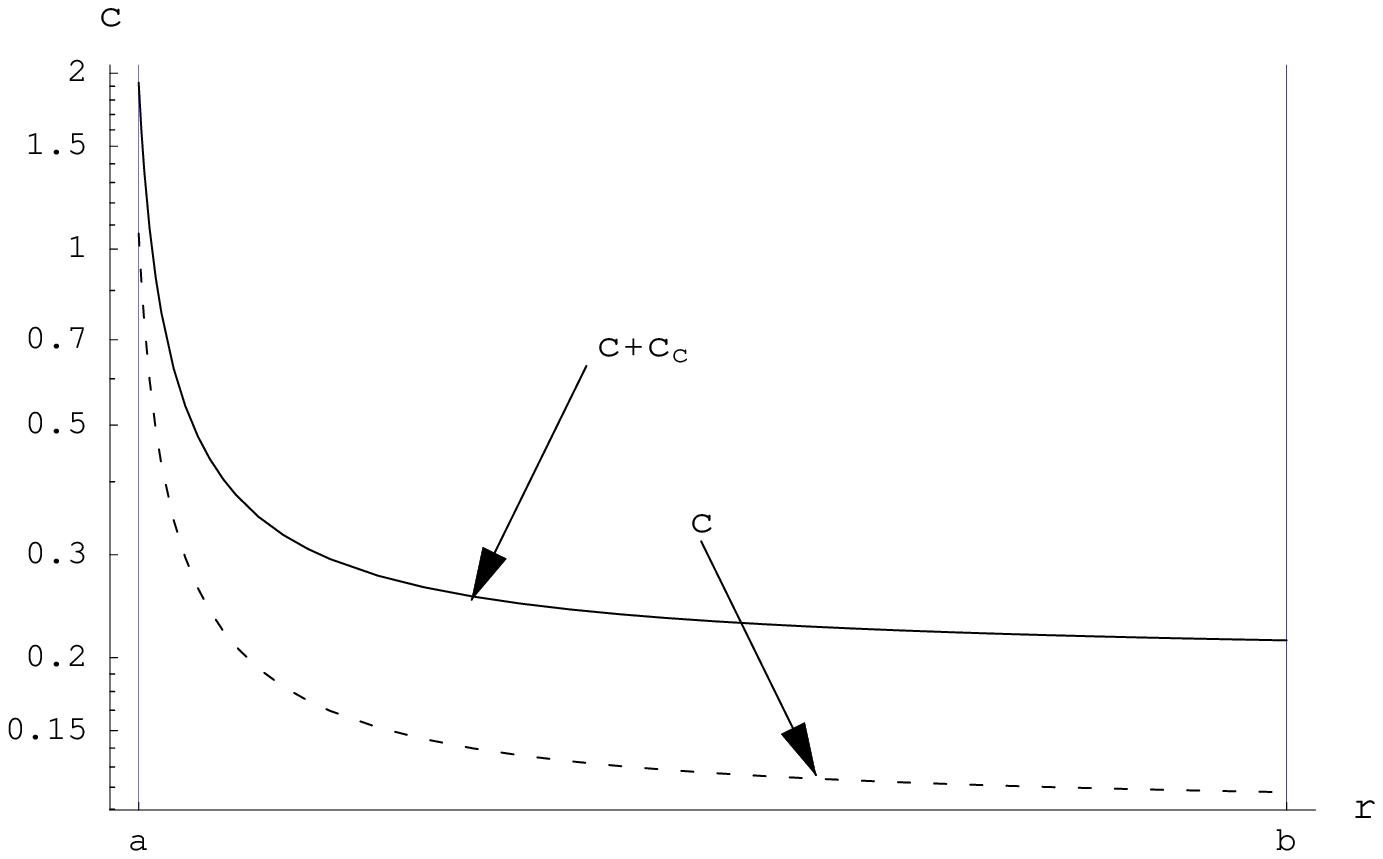}}
Figure 2: Space concentration of the free component $c(r)$ and
total diffusion component $c(r)+c_c(r)$\,.\\
\end{center}

In the solid body physics the role of
the reversible reaction is undertaken by the
process of reversible transitions of impurity
atoms between interstitial and substitution positions. The process of
this interchange can be represented as the chemical reaction with vacancy:

\begin{equation}
I + V \rightleftharpoons S
\end{equation}
($I$ and $S$ standing for interstitial and substitutional impurities,
$V$ for vacancies). Thus vacansies can be formally  understood as
"impurity carriers" like $Hb$ complexes. Another
mechanism is so called Watkins mechanism which also involves the
interchange of fast and slow species, but by the intermittency of
eigen interstitial atoms of the matrix. It is known
(e.g., \cite{Loualiche}) that such
reversible reactions are crucial for understanding the impurity
redistribution in the depth of a crystal.
In order to handle analytically the impurity profile, the authors
of \cite{Loualiche} used similar assumptions of rapid chemical
equilibrium establishing
compared to the ordinary diffusion.
In conclusion on Fig.2 we show the approximate solution for space
oxygen concentration in a spherical slab (between inner $a$ and
outer $b$ boundary). Shown are both diffusion $c(r)$ component and
the value $c+c_c$, that is total oxygen concentration.

\end{multicols}

\end{document}